# Effects of the generalized uncertainty principle on the thermal properties of Kemmer oscillator

Bing-Qian Wang[1,2], Zheng-Wen Long[2], Chao-Yun Long[2] and Shu-Rui Wu[2]

[1] College of mathematics and statistics, Guizhou University, Guiyang 550025, China

[2] College of Physics, Guizhou University, Guiyang 550025, China

A series of aspects of the quantum gravity predict a modification in the Heisenberg uncertainty principle to the generalized uncertainty principle (GUP). In the present work, using the momentum space representation, we study the behavior of the Kemmer oscillator in the context of the GUP. The wave function, the probability densities, and the energy spectrum are obtained analytically. Furthermore, the thermodynamic properties of the system are investigated via numerical method and the influence of GUP on thermodynamic functions is also discussed.

## 1.Introduction

The Kemmer equation introduced by Kemmer [1] is a two-body Dirac-like equation which involves matrices obeying a different scheme of commutation rules [1-3]. Historically, with the development of the equivalence of the Kemmer approach to the Klein–Gordon (KG) and Proca descriptions in on-shell situations, this equation tended to unpopular until it began to be investigated in several situations involving the breaking of symmetries and hadronic process, showing that in a way Kemmer and KG theories can give different results [4-5]. For these reasons, there has been a growing interest in studying the Kemmer equation [6-14].

On the other hand, it is well known that the generalizations of the uncertainty principle of quantum mechanics is probably the most challenging and interesting problem in theoretic physics. One of the main lines of investigation focuses on understanding how the Heisenberg Uncertainty Principle (HUP) should be modified once gravity is taken into account, thus various typic approaches such as string theory, loop quantum gravity, deformed special relativity and studies of black hole physics [15-22] are involved in the arguments. One of the well known formalism of the GUP is given by[23-26]

$$\Delta x_i \Delta p_i \geq \frac{\hbar}{2}[1 + \beta((\Delta p)^2 + <p>^2) + 2\beta(\Delta p_i^2 + <p_i>^2)], \tag{1}$$

where

$\beta = \beta_0/(M_p c)^2 = (l_p^2/\hbar^2)\beta_0$, $p^2 = \sum_i p_i p_i$, $M_p$ is the Planck mass and $M_p c^2$ corresponds to the Planck energy. In addition, inequality (1) is equivalent to the following modified Heisenberg algebra[19]

$$[x_i, p_j] = i\hbar(\delta_{ij} + \beta \delta_{ij} p^2 + 2\beta p_i p_j), \qquad (2)$$

via the Jacobi identity, this form ensures that $[x_i, x_j] = [p_i, p_j] = 0$.

Nowadays, various topics have been studied in connection with the GUP: Some aspects of the harmonic oscillator in the GUP scenario have been studied in [27-29]. The dynamics of a particle in a gravitational quantum well with GUP is studied in [30]. The quantum mechanical scattering problem for the Yukawa and the Coulomb potential has been studied in [31]. The implications of a generalized uncertainty relation on cosmology are discussed in many papers, for example see [32,33]. The effects of the modified uncertainty principle on the inflation parameters have been studied in [34]. The approaches and applications with generalized uncertainty principle are discussed in [35]. The Schrödinger equation in the presence of the GUP have been studied in [36-37]. The solution of the Dirac equation in the GUP scenario is solved in [38].

In spite of the great number of papers that have recently been published concerning the solutions and properties of the Kemmer equation [6-14], as far as we know, no one has reported on its thermal properties in the context of the GUP. In the present work, we are interested to study the effects of the generalized uncertainty principle on the thermal properties of the Kemmer oscillator. This work is organized as follows: in Section 2 we analyze the solutions of the Kemmer oscillator in the context of the GUP and obtain the energy spectrum as well as the corresponding probability density. Subsequently, on the basis of the above, the thermodynamic properties of the Kemmer oscillator are investigated by employing the zeta Epstein function in Section 3. Moreover, in order to have an intuitive understanding for the statistical properties of the physical system, we depict the numerical results of the thermodynamic functions with several figures and discuss the effect of the GUP parameter on thermal properties in Section 4. Finally, Section 5 is our conclusion.

**2. The Kemmer oscillator in the context of the GUP**

The Dirac-like relativistic Kemmer equation for spin-1 particles can be written as [1-3]

$$(\hat{\beta}^u \hat{p}_u - Mc)\psi_k = 0, \qquad (3)$$

where M is the total mass of two identical spin-1/2 particles. The 16×16 Kemmer matrices $\hat{\beta}^u$ (u=0, 1, 2 and 3) satisfy the relation

$$\hat{\beta}^u\hat{\beta}^v\hat{\beta}^\lambda + \hat{\beta}^\lambda\hat{\beta}^v\hat{\beta}^u = \hat{g}^{uv}\hat{\beta}^\lambda + \hat{g}^{\lambda v}\hat{\beta}^u, \tag{4}$$

with

$$\hat{\beta}^u = \hat{\gamma}^u \otimes \hat{I} + \hat{I} \otimes \hat{\gamma}^u, \tag{5}$$

where $\hat{I}$ is a 4×4 identity matrix, $\hat{\gamma}^u$ are the Dirac matrices, and $\otimes$ indicates a direct product.

Considering the configuration of the Dirac oscillator potential in the 1D case, the momentum operator $\hat{p}_x$ in the free Kemmer equation, could be substituted by $\hat{p}_x - iM\widehat{B}w\hat{x}$, where w is the oscillator frequency, and the operator $\widehat{B}$ is chosen as $\widehat{B} = \hat{\gamma}^0 \otimes \hat{\gamma}^0$, with $\widehat{B}^2 = \hat{I}$. So, the Kemmer equation with a Dirac oscillator interaction is

$$[(\hat{\gamma}^0 \otimes \hat{I} + \hat{I} \otimes \hat{\gamma}^0)E - c(\hat{\gamma}^0 \otimes \hat{\sigma}_x + \hat{\sigma}_x \otimes \hat{\gamma}^0)(\hat{p}_x - iM\widehat{B}w\hat{x}) - Mc^2\hat{\gamma}^0 \otimes \hat{\gamma}^0]\psi_k = 0, \tag{6}$$

where the Dirac $\hat{\gamma}$ matrices are substituted by Pauli $\hat{\sigma}$ matrices. The stationary state $\psi_k$ of equation (6) is four-component wave function of the Kemmer equation, which can be written as

$$\psi_k = \psi_D \otimes \psi_D = (\psi_1\ \psi_2\ \psi_3\ \psi_4)^T, \tag{7}$$

where $\Psi_D$ is the solution of the Dirac equation. Then substituting equation (7) into equation (6), we can easily obtain four linear algebraic equations

$$(2E - Mc^2)\psi_1 - c(\hat{p}_x + iMwx)\psi_2 - c(\hat{p}_x + iMwx)\psi_3 = 0,$$

$$-c(\hat{p}_x - iMwx)\psi_1 + Mc^2\psi_2 + c(\hat{p}_x - iMwx)\psi_4 = 0,$$

$$-c(\hat{p}_x - iMwx)\psi_1 + Mc^2\psi_3 + c(\hat{p}_x - iMwx)\psi_4 = 0,$$

$$c(\hat{p}_x + iMwx)\psi_2 + c(\hat{p}_x + iMwx)\psi_3 - (2E + Mc^2)\psi_4 = 0. \tag{8}$$

From these equations, we get the following results:

$$\psi_2 = \psi_3, \quad \psi_1 = \frac{2c}{2E - Mc^2}(\hat{p}_x + iMwx)\psi_2, \quad \psi_4 = \frac{2c}{2E + Mc^2}(\hat{p}_x + iMwx)\psi_2, \tag{9}$$

combination equation (8) and (9) gives

$$\left\{\hat{p}_x^2 + M^2w^2x^2 + \frac{Mc^2}{k_1 + k_2} - iMw[x, p_x]\right\}\psi_2 = 0, \tag{10}$$

where $k_1 = 2c^2/(Mc^2 - 2E)$, $k_2 = 2c^2/(Mc^2 + 2E)$. Here, it should be noted that in the momentum space representation we have $\hat{p} = p$ and $\hat{x} = i\hbar(1 + \beta p^2)\frac{\partial}{\partial p_x}$.

In addition, according to the eq. (2), the commutation relation between position vector and momentum vector satisfy

$$[x, p] = i\hbar(1 + \beta p^2). \tag{11}$$

Substituting (11) into (10) we have

$$\left[(1 + \beta p^2)^2 \frac{\partial^2}{\partial p^2} + 2\beta p(1 + \beta p^2)\frac{\partial}{\partial p} - \frac{(1 + Mw\hbar\beta)}{M^2w^2\hbar^2}p^2 - \frac{(\varepsilon + Mw\hbar)}{M^2w^2\hbar^2}\right]\psi_2(p) = 0, \tag{12}$$

where $\varepsilon = \frac{Mc^2}{k_1 + k_2}$.

With the aid of the variable q defined by $p \in (-\infty, +\infty) \to q \in \left(-\frac{\pi}{2Mw\hbar\sqrt{\beta}}, +\frac{\pi}{2Mw\hbar\sqrt{\beta}}\right)$
$q = \frac{\tan^{-1} p\sqrt{\beta}}{Mw\hbar\sqrt{\beta}}$, we can rewrite eq.(12) as

$$\left[\frac{\partial^2}{\partial q^2} - \frac{(1 + Mw\hbar\beta)}{\beta}\frac{\sin^2(Mw\hbar\sqrt{\beta}q)}{1 - \sin^2(Mw\hbar\sqrt{\beta}q)} - (Mw\hbar + \varepsilon)\right]\psi_2(q) = 0. \tag{13}$$

Here we introduce an auxiliary function $\psi_2(q) = [1 - \sin^2(Mw\hbar\sqrt{\beta}q)]^{\frac{\varsigma}{2}} f[\sin(Mw\hbar\sqrt{\beta}q)]$ and for the sake of simplification, let us assume that

$$\varsigma(\varsigma - 1) - \frac{(1+Mw\hbar\beta)}{M^2 w^2 \hbar^2 \beta^2} = 0. \tag{14}$$

Actually, this equation will lead to the following expression of $\varsigma$, i.e. $\varsigma_1 = \frac{1 + \sqrt{1 + \frac{4(1+Mw\hbar\beta)}{M^2 w^2 \hbar^2 \beta^2}}}{2}$ and $\varsigma_2 = \frac{1 - \sqrt{1 + \frac{4(1+Mw\hbar\beta)}{M^2 w^2 \hbar^2 \beta^2}}}{2}$. However, the second solution leads to the divergence of wave function, therefore, in the following, we select $\varsigma = \varsigma_1$. Thus the eq. (13) turns into

$$[1 - \sin^2(Mw\hbar\sqrt{\beta}q)]f''[\sin(Mw\hbar\sqrt{\beta}q)] - (2\varsigma + 1)\sin(Mw\hbar\sqrt{\beta}q)f'[\sin(Mw\hbar\sqrt{\beta}q)] +$$
$$\left[-\varsigma - \frac{(\varepsilon + Mw\hbar)}{M^2 w^2 \hbar^2 \beta}\right] f[\sin(Mw\hbar\sqrt{\beta}q)] = 0. \tag{15}$$

Moreover, the polynomial solution to eq. (15) is obtained by demanding the following condition:

$$-\varsigma - \frac{\varepsilon + Mw\hbar}{M^2 w^2 \hbar^2 \beta} = n(n + 2\varsigma) \tag{16}$$

with n a non-negative integer.

Then eq. (15) can be rewritten as

$$[1 - \sin^2(Mw\hbar\sqrt{\beta}q)]f''[\sin(Mw\hbar\sqrt{\beta}q)] - (2\varsigma + 1)\sin(Mw\hbar\sqrt{\beta}q)f'[\sin(Mw\hbar\sqrt{\beta}q)] +$$
$$n(n + 2\varsigma)f[\sin(Mw\hbar\sqrt{\beta}q)] = 0. \tag{17}$$

Obviously, its solution can be expressed in terms of Gegenbauer's polynomials as $f[\sin(Mw\hbar\sqrt{\beta}q)] = N C_n^{\varsigma}[\sin(Mw\hbar\sqrt{\beta}q)]$, with N is a normalization constant. Then the momentum eigenfunction of Kemmer oscillator in the context of the GUP are given by

$$\psi_2(p) = N[1 - \sin^2(Mw\hbar\sqrt{\beta}q)]^{\frac{\varsigma}{2}} C_n^\varsigma[\sin(Mw\hbar\sqrt{\beta}q)] = N\left(\frac{1}{1+\beta p^2}\right)^{\frac{\varsigma}{2}} C_n^\varsigma\left(\frac{p\sqrt{\beta}}{\sqrt{1+\beta p^2}}\right). \tag{18}$$

Besides, by using the following property of Gegenbauer's polynomials [39]

$$\frac{dC_n^\varsigma[\sin(Mw\hbar\sqrt{\beta}q)]}{d\sin(Mw\hbar\sqrt{\beta}q)} = 2\varsigma C_{n-1}^{\varsigma+1}[\sin(Mw\hbar\sqrt{\beta}q)], \tag{19}$$

we finally obtain

$$\psi_1(p) = \frac{2c}{2E - Mc^2}\left[p_x - \hbar Mw(1+\beta p^2)\frac{\partial}{\partial p_x}\right]\psi_2(p) = \frac{2c}{2E - Mc^2} N\left(\frac{1}{1+\beta p^2}\right)^{\frac{\varsigma}{2}}$$

$$\left\{(p + \hbar Mw\varsigma\beta p)C_n^\varsigma\left(\frac{p\sqrt{\beta}}{\sqrt{1+\beta p^2}}\right) - 2\varsigma\hbar Mw(1+\beta p^2)C_{n-1}^{\varsigma+1}\left(\frac{p\sqrt{\beta}}{\sqrt{1+\beta p^2}}\right)\left[\sqrt{\frac{\beta}{1+\beta p^2}} - \beta^{\frac{3}{2}}p^2(1+\beta p^2)^{\frac{-3}{2}}\right]\right\},$$

$$\psi_2(p) = \psi_3(p),$$

$$\psi_4(p) = \frac{2c}{2E + Mc^2}\left[p_x - \hbar Mw(1+\beta p^2)\frac{\partial}{\partial p_x}\right]\psi_2(p) = \frac{2c}{2E + Mc^2} N\left(\frac{1}{1+\beta p^2}\right)^{\frac{\varsigma}{2}}$$

$$\left\{(p + \hbar Mw\varsigma\beta p)C_n^\varsigma\left(\frac{p\sqrt{\beta}}{\sqrt{1+\beta p^2}}\right) - 2\varsigma\hbar Mw(1+\beta p^2)C_{n-1}^{\varsigma+1}\left(\frac{p\sqrt{\beta}}{\sqrt{1+\beta p^2}}\right)\left[\sqrt{\frac{\beta}{1+\beta p^2}} - \beta^{\frac{3}{2}}p^2(1+\beta p^2)^{\frac{-3}{2}}\right]\right\}.$$

$$\tag{20}$$

Therefore, the wave function of the system can be written as

$$\psi_k = N\begin{pmatrix} \frac{2c}{2E-Mc^2}(\hat{p}_x + iMwx) \\ 1 \\ 1 \\ \frac{2c}{2E+Mc^2}(\hat{p}_x + iMwx) \end{pmatrix} (1+\beta p^2)^{\frac{-\varsigma}{2}} C_n^\varsigma\left(\frac{p\sqrt{\beta}}{\sqrt{1+\beta p^2}}\right). \tag{21}$$

At this stage, we determine the normalization constant N by demanding the following normalization condition:

$$(\psi_k, \psi_k) = \int_{-\infty}^{+\infty} \frac{1}{1+\beta p^2} \psi_k^+ (\hat{\gamma}^0 \otimes \hat{\gamma}^0)\psi_k \, dp = 1, \tag{22}$$

and according to the property of Jacobi polynomial $\int_{-1}^{+1} dy(1-y^2)^{\lambda-\frac{1}{2}}[C_n^\lambda(y)]^2 = \frac{\pi 2^{1-2\lambda}\Gamma(2\lambda+n)}{n!(n+\lambda)[\Gamma(\lambda)]^2}$, we have

$$N = \beta^{\frac{1}{4}}\pi^{-\frac{1}{2}}2^{\varsigma-1}\left\{\frac{-\Gamma(2\varsigma+n)}{n!(n+\varsigma)[\Gamma(\varsigma)]^2} + \frac{2c^2(8E^2+2M^2c^4)}{(2E-Mc^2)^2(2E+Mc^2)^2}\left[\left(\frac{1}{\beta}-4\varsigma Mw\hbar\right)(1+Mw\hbar\varsigma\beta)\frac{\Gamma(2\varsigma+n)}{n!(n+\varsigma)[\Gamma(\varsigma)]^2} + \frac{(\varsigma Mw\hbar)^2\beta\Gamma(2\varsigma+n+2)}{(n+1)!(n+\varsigma+1)[\Gamma(\varsigma+1)]^2}\right]\right\}^{-\frac{1}{2}}, \qquad (23)$$

then the corresponding probability density of every component can be expressed as

$$P'_i = \left|\int_{-\infty}^{+\infty}\psi_i^\dagger(\hat{\gamma}^0\otimes\hat{\gamma}^0)\psi_i dp\right|, \text{i=1,2,3,4}. \qquad (24)$$

Furthermore, we derive the energy spectrum from (16) which leads to

$$E_n^2 = c^2M^2w^2\hbar^2\beta n^2 + (2Mw\hbar c^2\sqrt{1+Mw\hbar\beta} + c^2M^2w^2\hbar^2\beta)n + \frac{c^2M^2w^2\hbar^2\beta}{2} + \frac{c^4M^2}{4} + Mw\hbar c^2 + Mw\hbar c^2\sqrt{1+Mw\hbar\beta}. \text{ n=0,1,2…} \qquad (25)$$

In view of the obscurity and complexity of eq. (25), we decide to depict the numerical results aiming to show the effect of the GUP in the energy spectra. In figure 1, the energy spectra E versus the principal quantum number for different values of the GUP parameter are plotted. Positive and negative energy levels correspond to the case of particle and antiparticle, respectively. It shows that for the same principal quantum number, the energy E increases monotonically with the increase of the GUP parameter. The effect of the GUP parameter on the energy levels is observable, where $\beta = 0$ corresponding to the case of the normal quantum mechanics, and this result is rigorously consistent with the reference 6.

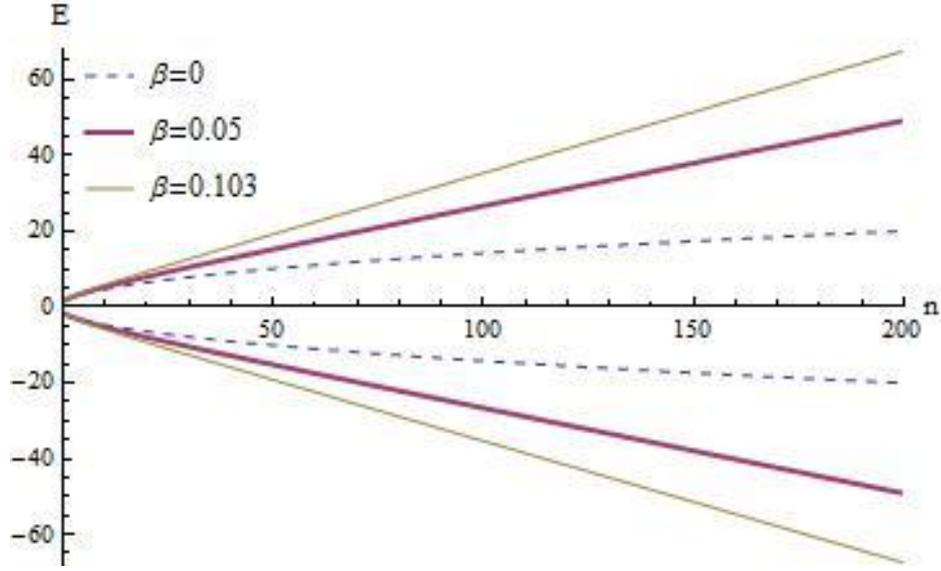

Figure 1. The profile of the energy spectra $E$ of the Kemmer oscillator in the context of GUP versus the principal quantum number $n$ for different values of $\beta$ ($M = \hbar = w = 1$).

### 3. The thermal function of Kemmer oscillator under the influence of GUP

As we know, all thermodynamic quantities can be obtained from the partition function $Z$, therefore, in the following work, the partition function of the system is calculated firstly. We start with the following eigenvalues of Kemmer oscillator in the context of GUP as

$$E_n = \sqrt{\kappa n^2 + \eta n + 1}\sqrt{\frac{M^2 c^2 w^2 \hbar^2 \beta}{2} + \frac{c^4 M^2}{4} + c^2 M w\hbar + c^2 M w\hbar\sqrt{1 + Mw\hbar\beta}}, \qquad (26)$$

where $\kappa = \dfrac{Mw^2\hbar^2\beta}{\frac{Mw^2\hbar^2\beta}{2} + \frac{c^2 M}{4} + w\hbar + w\hbar\sqrt{1 + Mw\hbar\beta}}$ and $\eta = \dfrac{2w\hbar\sqrt{1+Mw\hbar\beta} + Mw^2\hbar^2\beta}{\frac{Mw^2\hbar^2\beta}{2} + \frac{c^2 M}{4} + w\hbar + w\hbar\sqrt{1+Mw\hbar\beta}}$.

Given the energy spectrum, we can define the partition function via

$$Z = \sum_n e^{-\tilde{\beta} E_n}, \qquad (27)$$

where $\tilde{\beta} = \dfrac{1}{k_B T}$ with $k_B$ is the Boltzmann constant. In this case, eq. (27) can be written as

$$Z = \sum_n e^{-\frac{1}{T}\sqrt{\kappa n^2 + \eta n + 1}}, \qquad (28)$$

with $\text{I} = \frac{Mwc\hbar\tau}{\sqrt{\kappa}}$. Now we set $\tau = \frac{k_B T}{Mc^2}$, $t = \frac{1}{\text{I}}\sqrt{\kappa n^2 + \eta n + 1}$, and by using the following relation [40]

$$e^{-t} = \frac{1}{2\pi i}\int_c^\infty ds\, t^{-s}\Gamma(s), \qquad (29)$$

then the sum can be expressed as

$$\sum_n e^{-\frac{1}{\text{I}}\sqrt{\kappa n^2+\eta n+1}} = \frac{1}{2\pi i}\int_c^\infty ds\left(\frac{1}{\text{I}}\right)^{-s}\sum_n\{\kappa n^2+\eta n+1\}^{-\frac{s}{2}}\Gamma(s) = \frac{1}{2\pi i}\int_c^\infty ds\left(\frac{1}{\text{I}}\right)^{-s} J(s)\Gamma(s), (30)$$

where $J(s) = \sum_n \frac{1}{Q(1,n)^{\frac{s}{2}}}$ and $\Gamma(s)$ are the Euler and one-dimensional Epstein zeta function respectively, with $Q(1,n) = \kappa n^2 + \eta n + 1$. For the sake of convenience, we set $x = \frac{\eta}{2}$ and $y = \frac{\sqrt{4\kappa-\eta^2}}{2}$, then the expression of $J(s)$ can be rewritten as

$$J(s) = 2\kappa^{-\frac{s}{2}}\zeta(s) + \frac{2\kappa^{-\frac{s}{2}} y^{1-s}\sqrt{\pi}}{\Gamma\left(\frac{s}{2}\right)}\zeta(s-1)\Gamma\left(\frac{s}{2}-\frac{1}{2}\right) + \frac{2\kappa^{-\frac{s}{2}} y^{\left(\frac{1}{2}-\frac{s}{2}\right)}\pi^{\frac{s}{2}}}{\Gamma\left(\frac{s}{2}\right)}H\left(\frac{s}{2}\right). \qquad (31)$$

Now, the final partition function turns into

$$Z = \frac{\int_c^\infty ds\left(\frac{1}{\text{I}}\right)^{-s} 2\kappa^{-\frac{s}{2}}\zeta(s)\Gamma(s)}{2\pi i} + \frac{1}{2\pi i}\int_c^\infty ds\left(\frac{1}{\text{I}}\right)^{-s}\frac{2\kappa^{-\frac{s}{2}} y^{1-s}\sqrt{\pi}}{\Gamma\left(\frac{s}{2}\right)}\zeta(s-1)\Gamma\left(\frac{s}{2}-\frac{1}{2}\right)\Gamma(s)$$

$$+ \frac{1}{2\pi i}\int_c^\infty ds\left(\frac{1}{\text{I}}\right)^{-s}\frac{2\kappa^{-\frac{s}{2}} y^{\frac{1}{2}-\frac{s}{2}}\pi^{\frac{s}{2}}}{\Gamma\left(\frac{s}{2}\right)}H\left(\frac{s}{2}\right)\Gamma(s), \qquad (32)$$

Next, by applying the residues theorem, we have

$$Z = 2\zeta(0) + \frac{2}{\sqrt{\kappa}}\{\zeta(1)+\zeta(0)\}\text{I} + \frac{2\pi \text{I}^2}{\kappa y}. \qquad (33)$$

In addition, it should be noted that the last integral in equation (32) goes to the zero because of the following relation

$$\frac{1}{\Gamma(s)} = se^{rs} \prod_{n=1}^{\infty}\left\{\left(1+\frac{x}{n}\right)e^{-\frac{x}{n}}\right\}, \tag{34}$$

where r is Euler's constant expressed as $r = \lim_{n\to\infty}\left(\sum_{k=1}^{n}\frac{1}{k} - \log(n)\right)$, thus the final partition function of Kemmer oscillator in the context of GUP becomes

$$Z(\mathcal{T},\kappa) = \frac{2\pi}{\kappa\sqrt{\kappa-1}}\mathcal{T}^2 + \frac{1}{\sqrt{\kappa}}\mathcal{T} - 1. \tag{35}$$

Then the thermodynamic properties of the physical system, such as free energy, mean energy, specific heat, and entropy, can be calculated from the following expressions

$$F = -\mathcal{T}\ln(Z),$$

$$U = \mathcal{T}^2 \frac{\partial \ln(Z)}{\partial \mathcal{T}},$$

$$C = 2\mathcal{T}\frac{\partial \ln(Z)}{\partial \mathcal{T}} + \mathcal{T}^2 \frac{\partial^2 \ln(Z)}{\partial \mathcal{T}^2},$$

$$S = \ln(Z) + \mathcal{T}\frac{\partial \ln(Z)}{\partial \mathcal{T}}. \tag{36}$$

In order to perform our analysis on the thermodynamics of the Kemmer oscillator, we will restrict ourselves to stationary states of positive energy. And from (36), we predict that the thermodynamic functions will be very complicated. In this case, in order to have an intuitive understanding for thermodynamic properties of the Kemmer oscillator in the context of the GUP, in the following, we briefly depict our numerical results on the evaluation of the thermodynamic functions, i.e. free energy, mean energy, specific heat, and entropy, via the numerical partition function $Z$.

**4. Results and Discussions**

Before beginning, it should be noted that all profiles of the thermodynamic quantities as a function of dimensionless temperature variable $\tau$ for different values of $\kappa$, i.e. $\kappa =$

1.5, 2.3, 3.4 are plotted in Figures 2–6, and in these figures, the natural unit ($\hbar = c = M = 1$) is employed.

From the result shown in Figure 2, it shows that the partition function $Z$ increases monotonically with dimensionless variable $\tau$, and for a fixed value of $\tau$, the partition function decreases with the increase of the deformed parameter $\kappa$. In Figure 3, the free energy $F$ is shown, we see that the free energy decreases with $\kappa$ growing, and for a fixed value of $\tau$, the profile of the curves decreases monotonically with the temperature for both cases. We plot the mean energy U versus $\tau$ for different values of the deformed parameter $\kappa$ in Figure 4, it also shows that for a fixed $\tau$ the mean energy decreases when $\kappa$ grows. The profile of heat capacity C as a function of $\tau$ for different values of $\kappa$ is depicted in Figure 5. We show that the heat capacity increases for increasing $\tau$ at first and then remain invariant for a same value with $\tau$ growing, it means that the effect of the GUP on the heat capacity can be negligible at high temperature. The curves of the numerical entropy versus $\tau$ for different values of $\kappa$ is plotted in Figure 6. It shows that the tendency of entropy rapidly increases at first for increasing $\tau$ and then slowly grows for large $\tau$ values, and for a fixed value of $\tau$, the entropy of the system decreases when the deformed parameter $\kappa$ grows.

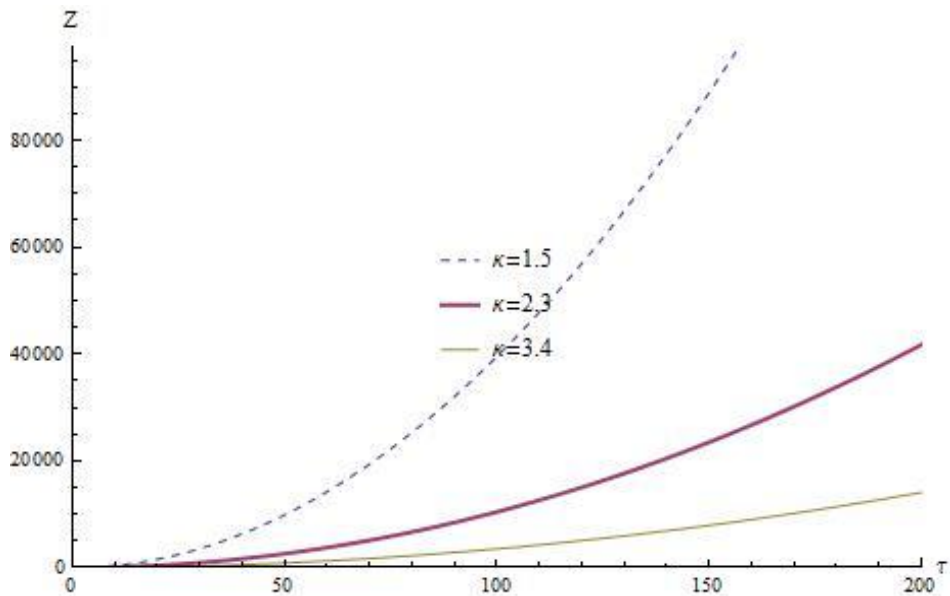

Figure 2.The partition function $Z$ of the Kemmer oscillator in the context of the GUP

as a function of $\tau$ for different values of $\kappa$.

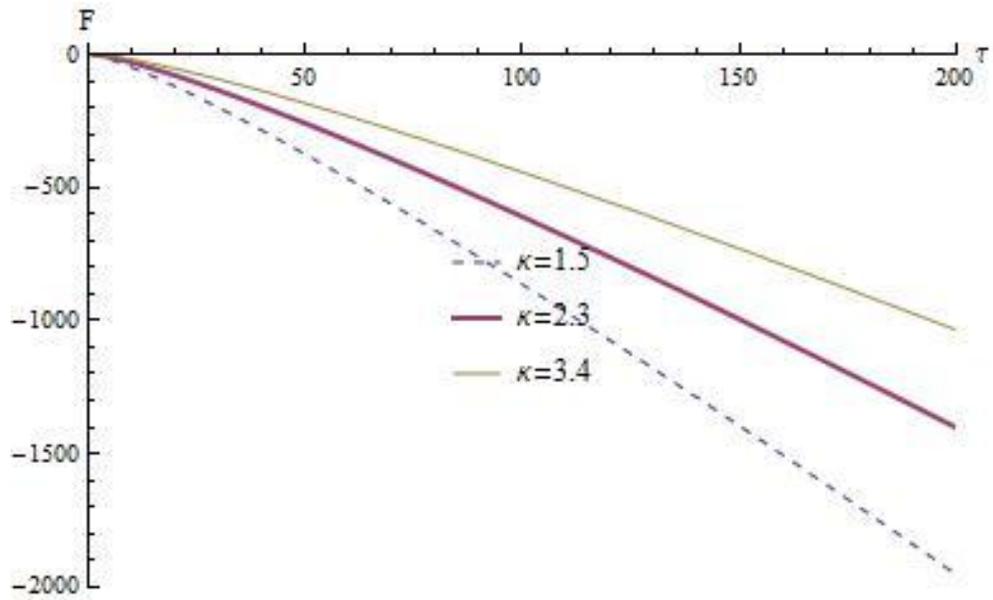

Figure 3. The free energy F of the Kemmer oscillator in the context of the GUP

as a function of $\tau$ for different values of $\kappa$.

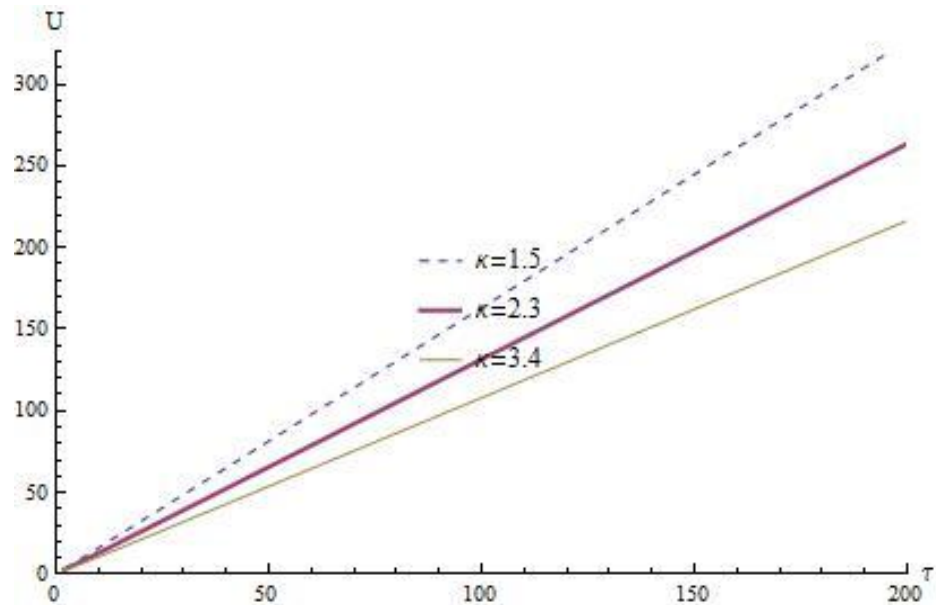

Figure 4. The mean energy U of the Kemmer oscillator in the context of the GUP

as a function of $\tau$ for different values of $\kappa$.

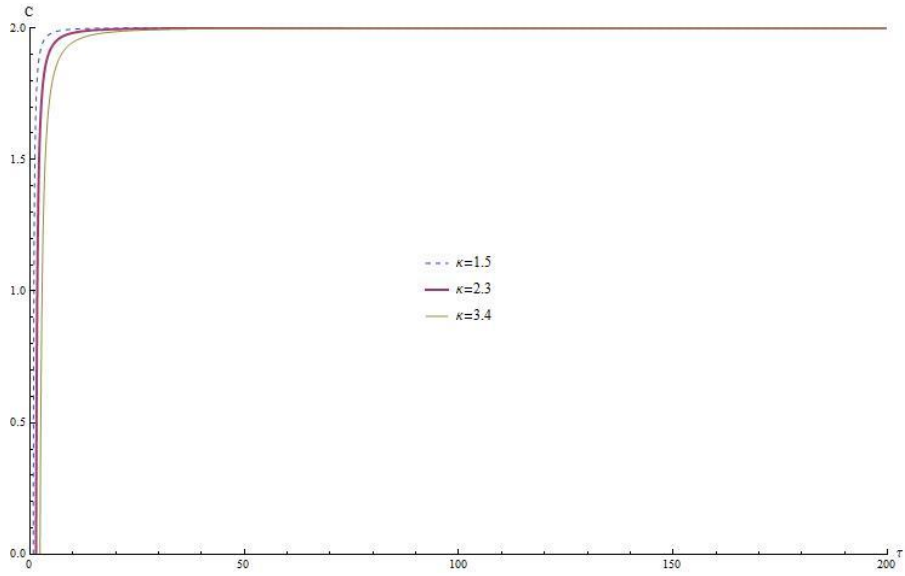

Figure 5.The specific heat C of the Kemmer oscillator in the context of the GUP

as a function of $\tau$ for different values of κ.

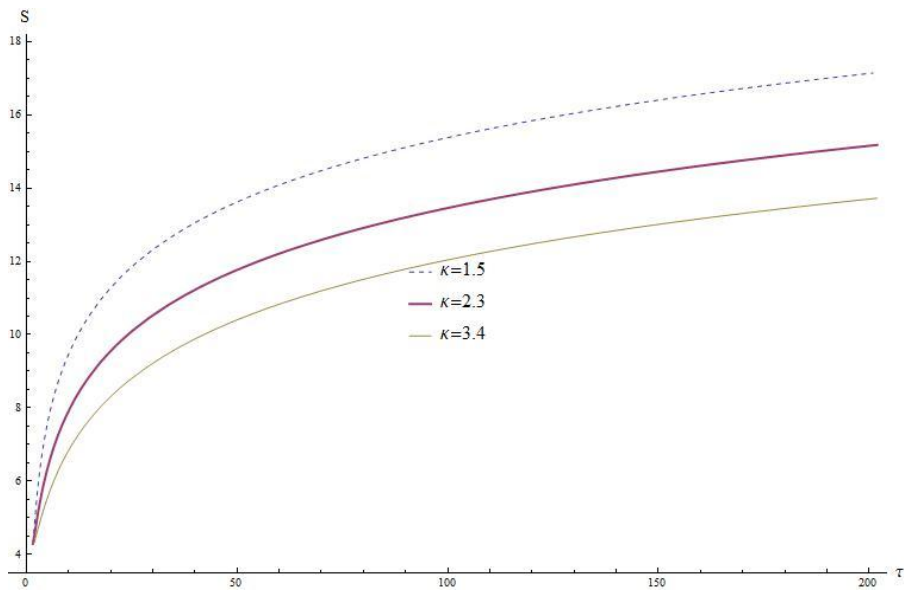

Figure 6.The entropy S of the Kemmer oscillator in the context of the GUP

as a function of $\tau$ for different values of κ.

## 5. Conclusions

This paper was devoted to study of the thermodynamic properties of the Kemmer oscillator in the context of the GUP. We first analyzed the Kemmer oscillator in the context of the GUP and obtained the wave function, the corresponding probability density of every component as well as the energy spectrum by employing the Gegenbauer polynomial. Subsequently, we investigated the thermodynamic properties of the system by employing the Epstein zeta function and depicted our numerical results for the corresponding thermodynamic functions through the associated partition function $Z$, and the effect of the GUP parameter on thermodynamic properties was reported. It shows that the wave function, the probability densities, the energy spectrum and the corresponding thermodynamic functions of Kemmer oscillator depend explicitly on the deformed parameter $\beta$ which characterizes the effects of the GUP on this considered physics system.

**Conflict of Interests**

The authors declare that there is no conflict of interests regarding the publication of this paper.

**Acknowledgment**

This work is supported by the National Natural Science Foundation of China (Grant Nos. 11465006, 11565009).

**References**

[1] N. Kemmer, "Quantum theory of einstein-bose particles and nuclear interaction," Proceedings of the Royal Society of London Series A: Mathematical and Physical Sciences, vol. 166, no. 924, pp. 127–153, 1938.

[2] R. J. Duffin, "On the characteristic matrices of covariant systems," Physical Review, vol. 54, no. 12, p. 1114, 1938.

[3] G. Petiau, "University of Paris thesis," Acad´emie Royale De Belgique. Classe Des Sciences. M émoires. Collection, vol. 16, no. 2, p. 1, 1936.

[4] Friedman E and Kalbermann G, "Kemmer-Duffin-Petiau equation for pionic atoms and anomalous strong interaction effects", Phys. Rev. C, vol. 34, no. 4, pp. 2244-2248, 1986.

[5] Casana R, Fainberg V Ya, Pimentel B M, and Valverde J S, "Bose–Einstein condensation and free DKP field" Phys. Lett. A, vol. 316, no. 11, pp. 33-43, 2003.

[6] Boumali A, "One-dimensional thermal properties of the Kemmer oscillator", Phys. Scr. vol. 76, no. 5, pp. 669-673, 2007.

[7] Boumali A and Chetouani L, "Exact solutions of the Kemmer equation for a Dirac oscillator", Phys. Lett. A, vol. 346, no. 8, pp. 261-268, 2005.

[8] Boutabia B and Boudjedaa T, "Solution of DKP equation in Woods–Saxon potential" Phys. Lett. A, vol. 338, no. 11, pp. 97-107, 2005.

[9]Chetouani L, Merad M, Boudjedaa T, and Lecheheb A, "Solution of Duffin–Kemmer–Petiau Equation for the Step Potential" , Int. J. Theor. Phys. vol. 43, no. 13, pp. 1147-1159, 2004.

[10] Boumali A, "The spin 0 particle in an Aharanov-Bohm potention" Can. J. Phys. vol. 82, no. 8, pp. 67-74, 2004.

[11] Fernandes M C B, Santana A E, and Vianna J D M, "Galilean Duffin–Kemmer–Petiau algebra and symplectic structure",  J. Phys. A: Math. Gen. vol. 36, no. 13, pp. 3841-3854, 2003.

[12] Ghose P, Samal M K, and Datta A, "Klein paradox for bosons", Phys. Lett. A, vol. 315, no. 5, pp. 23-27, 2003.

[13] Bolivar A O, "Classical limit of bosons in phase space",  Physica A, vol. 315, no. 15, pp. 601-615, 2002.

[14] Lunardi J T, Pimentel B M, Teixeiri R G, and Valverde J S, "Remarks on Duffin–Kemmer–Petiau theory and gauge invariance", Phys. Lett. A, vol. 268, no. 9, pp. 165-173, 2000.

[15] D. J. Gross, P.F. Mende, "The high-energy behavior of string scattering amplitudes", Phys. Lett. B, vol. 197, no. 6, pp. 129-134, 1987.

[16] D. Amati, M. Ciafaloni, and G. Veneziano, "Can spacetime be probed below the string size?", Phys. Lett. B, vol. 216, no.7, pp. 41-47, 1989.

[17] K .Konishi, G. Paffuti, and P. Provero, "Minimum physical length and the generalized uncertainty principle in string theory", Phys. Lett. B, vol. 234, no. 9, pp. 276-284, 1990.

[18] M. Maggiore, "A generalized uncertainty principle in quantum gravity", Phys. Lett. B, vol. 304, no. 5, pp. 65-69, 1993.

[19] A. Kempf, G. Mangano, and R.B. Mann, "Hilbert Space Representation of the Minimal Length Uncertainty Relation", Phys. Rev. D, vol. 52, no. 2, pp. 1108-1109, 1995.

[20] M. Bojowald and A. Kempf, Phys. Rev. D, vol. 86, Article ID 085017, 2012.

[21] F. Scardigli, "Generalized uncertainty principle in quantum gravity from micro-black hole gedanken experiment" , Phys. Lett. B, vol. 452, no. 6, pp. 39-44, 1999.

[22] F. Scardigli and R. Casadio, "Generalized uncertainty principle, extra dimensions and holography", Class. Quantum Grav. vol. 20, no. 18, pp. 3915-3914, 2003.

[23] M. Maggiore, "Quantum groups, gravity, and the generalized uncertainty principle", Phys. Rev. D, vol. 49, no. 6, pp. 5182-5187, 1994.

[24] Ahmed Farag Ali, Saurya Das, and Elias C. Vagenas, Rev.D, vol. 84, Article ID 044013, 2011.

[25] S. Haouat and K. Nouicer, "Influence of a minimal length on the creation of scalar particles",   Phys. Rev. D 89, Article ID 105030, 2014.

[26] T. L. Antonacci Oakes, R. O. Francisco, J. C. Fabris, and J. A. Nogueira, "Ground state of the hydrogen atom via Dirac equation in a minimal-length scenario", Eur. Phys. J.C, vol. 73, no. 6, pp. 2495-2500, 2013.

[27] Nouicer K, "Path integral for the harmonic oscillator in one dimension with nonzero minimum position uncertainty", Phys. Lett. A, vol. 354, no. 6, pp. 399-405, 2006.

[28] Nozari K and Azizi T, "Gravitational induced uncertainty and dynamics of harmonic oscillator", Gen. Rel. Grav. vol. 38, no. 7, pp. 325-331, 2006.

[29 Bing-Qian Wang, Zheng-Wen Long, Chao-Yun Long, and Shu-Rui Wu, "(2+ 1)-Dimensional Duffin-Kemmer-Petiau Oscillator under a Magnetic Field in the Presence of a Minimal Length in the Noncommutative Space" Advances in High Energy Physics, vol. 2017, Article ID 2843020, 2017.

[30] Brau F and Buisseret F, "Minimal Length Uncertainty Relation and gravitational quantum well", Phys. Rev. D, vol.74, Article ID 036002, 2006.

[31] Stesko M M and Tkachuk V M, "Scattering problem in deformed space with minimal length", Phys. Rev. A 76, Article ID 012707, 2007.

[32] Battisti M V and Montani G, "Quantum dynamics of the Taub universe in a generalized uncertainty principle framework", Phys. Rev. D, 77, Article ID 023518, 2008.

[33] Battisti M V and Montani G, "The Big-Bang singularity in the framework of a Generalized Uncertainty Principle", Phys. Lett. B, vol. 656, no. 6, pp. 96-101, 2007.

[34] Momotaj Ara, Md Moniruzzaman and S B Faruque, "Exact solution of the Dirac equation with a linear potential under the influence of the generalized uncertainty principle", Phys. Scr. vol. 82, Article ID 035005, 2010.

[35] B Majumder, "Effects of the modified uncertainty principle on the inflation parameters", Physics Letters B, vol. 709, no. 4, pp. 133-136, 2012.

[36] WANG Lun-Zhou, LONG Chao-Yun, and LONG Zheng-Wen, "Quantization of Space in the Presence of a Minimal Length", Commun. Theor. Phys. vol. 63, no. 6, pp. 709-714, 2015.

[37] Bing-Qian Wang, Chao-Yun Longa, Zheng-Wen Long, and Ting Xu, "Solutions of the Schrödinger equation under topological defects space-times and generalized uncertainty principle", Eur. Phys. J. Plus, vol. 131, no. 7, pp. 378-384, 2016.

[38] B Majumder, "Effects of the modified uncertainty principle on the inflation parameters", Physics Letters B, 709, vol. 4, pp. 133-136, 2012.

[39] Gradshteyn I S and Ryzhik I M, Tables of Integrals, Series and Products, New York, Academic 1980.

[40] A. Boumali, "The one-dimensional thermal properties for the relativistic harmonic oscillators", arXiv:1409.6205v1.